# Studying the evolution of exchange and dipolar interactions and spin correlation in vdW magnet $CrCl_3$ across magnetic phases

Prem B. Karki,[1] Ayodimeji E. Aregbesola,[1] Karishma Prasad,[2] Jonathan Skeels,[1] Avani Inamdar[1], Pooja Kalal[1], Jian Wang,[2] Kapildeb Ambal[1]*

[1]Department of Mathematics, Statistics, and Physics, Wichita State University, 1845 Fairmount Street, Wichita, KS 67260, USA

[2]Department of Chemistry and Biochemistry, Wichita State University, 1845 Fairmount Street, Wichita, KS 67260, USA

* Corresponding author: Kapildeb.ambal@wichita.edu

Understanding quantum interactions in van der Waals magnets are critical to realizing their potential for next-generation magnetic, electronic, and spintronic applications. In this study, we investigated the evolution of exchange and dipole-dipole interactions in van der Waals magnet $CrCl_3$ using multifrequency broadband (1-16 GHz) magnetic resonance on a bulk single-crystal $CrCl_3$ sample. We found that the dipolar-broadened low-field resonance linewidth narrows due to exchange narrowing at higher resonance fields. Our findings show that the exchange interaction evolves exponentially as we lower the sample temperature from room temperature to 50 K. However, we found that the dipole-dipole interaction remains constant in the same temperature range. We have also detected the half-field magnetic resonance transition at room temperature, indicating a robust dipole-dipole interaction among spins. The measured inter-spin distance is 2.4(3) Å at room temperature. The evolution of dipole interactions, the observation of half-field magnetic resonance, and the strong exchange interaction indicate short-range ordering/correlation in $CrCl_3$ at temperatures far above the magnetic global ordering temperature. This study also demonstrates the significance and sensitivity of the multifrequency electron spin resonance technique for understanding quantum interactions and magnetic correlations in van der Waals magnetic materials.

## Introduction:

Two-dimensional van der Waals magnets have emerged as promising platforms for next-generation spintronic and quantum devices. Interactions between quantum spins influence the magnetic, thermal, and transport properties of 2D vdW magnetic materials through mechanisms like exchange interactions, dipole-dipole forces, and spin-orbit coupling. These interactions drive dynamics such as spin waves (magnons), topological ordering, and quantum spin liquid states. When spins interact, they often produce emergent phenomena that differ from the behavior of isolated spins. Therefore understanding quantum interactions in van der Waals magnets are critical to realizing their potential for next-generation magnetic, electronic, and spintronic applications.

Among them, transition-metal trihalides $CrX_3$ (X = Cl, Br, I) have emerged as strong contenders for highly tunable platforms in which external stimuli, such as strain or electric fields, can switch magnetic phases. While $CrBr_3$ and $CrI_3$ exhibit a strong out-of-plane Ising-type anisotropy [1], $CrCl_3$ is characterized by a weak easy-plane XY anisotropy, with spins preferring to settle within the honeycomb layers [2–4]. This unique configuration makes $CrCl_3$ an ideal platform for studying many phenomena [5,6], including layer-dependent tunneling magnetoresistance [7], pressure-driven magnetic switching [8], enhanced spin polarization under an applied magnetic field [9], and antiferromagnetic resonance in the gigahertz frequency range [7]. The physics behind the magnetic ordering in $CrCl_3$ is notably complex and controlled by the exchange and dipolar interactions among the spins and spin-orbit coupling (SOC) [10,11]. Previous studies have proposed the possibility of short-range magnetic order in $CrCl_3$ well above the ferromagnetic transition temperature [12]. The low magnetic transition temperatures of $CrCl_3$ provide an ideal platform for studying short-range magnetic order at temperatures well above the global ordering temperature. Previous studies using inelastic Neutron scattering and Raman spectroscopy have revealed magnon features that remain identifiable well above the global ordering temperature [13].

Electron magnetic resonance is a useful high-resolution (~μeV) spectroscopic technique that enables the study of local magnetic properties and magnetic phase transitions in magnetic systems. Multifrequency electron magnetic resonance [14–17], at both low and high magnetic fields, has opened a new, sensitive window to probe magnetic interactions and electron spin dynamics, whose influence on magnetic resonance spectra depends strongly on the applied magnetic field strength [16]. For example, it is well established that spin exchange leads to a narrowing of electron magnetic resonance signals, and that the exchange interaction can be estimated from the variation of the signal linewidth and/or line shape as a function of applied magnetic field strength [18–20]. Conventional multifrequency electron magnetic resonance experiments are carried out only at a few selected, reasonably high magnetic fields for which commercial resonators and microwave bridges are readily available, and for materials in which strong thermal spin polarization is observed. This is because spin polarization is measured in electron magnetic resonance experiments, and it is limited by the Boltzmann factor that determines thermal spin polarization, which decreases with decreasing Zeeman splitting (lower magnetic field) and increasing temperature. Previously, multifrequency electron magnetic resonance experiments on $CrCl_3$ were carried out on X-band (> 9 GHz or > 0.3 T) and higher frequencies [21,22]. To our knowledge, exchange narrowing has never been reported in $CrCl_3$, and we are the first to report the observation of exchange narrowing in $CrCl_3$ over a wide temperature range.

In this work, we study the evolution of exchange and magnetic dipole-dipole interactions in van der Waals magnet $CrCl_3$ using multifrequency broadband (1-16 GHz) electron magnetic resonance on a bulk single-crystal $CrCl_3$ sample from room temperature (RT) to 50 K. The exchange narrowing of the magnetic resonance linewidth at high resonance frequencies is evident from the pronounced reduction in resonance linewidth from low to high resonance frequencies. Using Anderson-Weiss theory [19], we extract the exchange and dipole-dipole interactions at a given temperature. We presented a detailed investigation of the temperature dependence of exchange and dipole-dipole interactions. We also measured half-field magnetic resonance ($\Delta m = \pm 2$) at room temperature, indicating robust dipole-dipole interactions. We further used the bootstrap method to extrapolate the temperature-dependent exchange and dipole interactions beyond our measurement range and predicted them in the ferromagnetic and antiferromagnetic (~2-14 K) domains. We measured the broadband ferromagnetic resonance at 14 K and extracted the exchange interaction, which agrees with the predicted value. Our predicted exchange interaction in the antiferromagnetic phases of $CrCl_3$ also agrees with literature values. We further investigated the short-range order in $CrCl_3$ by examining the temperature dependence of the resonance linewidth. From the measurement outcomes, we quantified the spin correlation length from room temperature down to 20 K, in agreement with the literature prediction.

## **Experimental Methods**:

Bulk $CrCl_3$ crystals were synthesized following the procedure reported elsewhere [23]. A brief description is also provided here. We grew single crystals of $CrCl_3$ using a self-transport technique starting from commercially available $CrCl_3$ powder (Thermo Fisher, 99.9%). Approximately 0.6 g of $CrCl_3$ was loaded into a 20 cm long quartz ampoule of inner diameter 17 mm. The ampoule was sealed under vacuum and placed horizontally in a tube furnace, with the starting material near the center (hot zone) of the furnace, while the opposite end extended into the cooler region near the furnace opening. The furnace was heated to 700 °C over 24 h and maintained at this temperature for an additional 24 h to ensure homogenization. Subsequently, a temperature gradient of 550–450 °C (hot-to-cold end) was established and maintained for several days to facilitate crystal growth via vapor transport. Large violet-colored $CrCl_3$ single crystals formed at the cooler end of the tube. To obtain larger crystals, we repeated the growth procedure under identical conditions. Millimeter-sized crystals of $CrCl_3$ were collected (Figure 1b) and stored in an Ar-filled glovebox. We verified the phase identification and purity of $CrCl_3$ by X-ray diffraction, and the powder X-ray diffraction results are presented in Figure S1 in the Supporting Information. We mechanically cleaved the bulk crystals to produce a clean, uniformly thick single-crystal $CrCl_3$ sample. We then transferred the selected sample onto a coplanar waveguide (CPW). We fixed the $CrCl_3$ sample on the waveguide using polyimide (Kapton) tape as shown in Figure 1a.

We mounted the CPW assembly on a measurement platform for broadband electron magnetic resonance measurements, which consists of an electromagnet, a modulation coil, an RF frequency source, a lock-in amplifier, a closed-cycle cryostat, and a computer with a data acquisition card (DAQ). We applied a static magnetic field along with a small sinusoidal modulation of 0.5 mT at ~77 Hz for lock-in detection. We applied microwave power of 15 dBm for the electron magnetic resonance excitation. The modulation parameters and microwave power were optimized to avoid magnetic modulation-induced linewidth broadening [24]. The signal was detected using a lock-in amplifier with a lock-in time constant of 0.3 s, and the dwell time per data point was 2 s. We digitized the data using a data acquisition card (DAQ) on the computer. We mounted the sample so that the applied static magnetic field was aligned with the in-plane (ab plane) direction. Broadband magnetic resonance was performed from 1–16 GHz, limited by the magnetic field strength of our laboratory electromagnet.

We also measured half-field magnetic resonance. For this purpose, we used a commercial X-band Bruker electron spin resonance (ESR) spectrometer. In this measurement, we attached a single-crystal $CrCl_3$ sample to a glass substrate. We then placed the sample inside the resonator so that the applied magnetic field aligned with the in-plane (ab-plane) direction. We measured full-field resonance to determine the resonance field, followed by half-field resonance measurement. In this measurement, we modulated the magnetic field at 6 kHz with an amplitude of 0.5 mT, limited by the instrument's capabilities. The resonator Q-value was about 3000, and the microwave power was 20 mW.

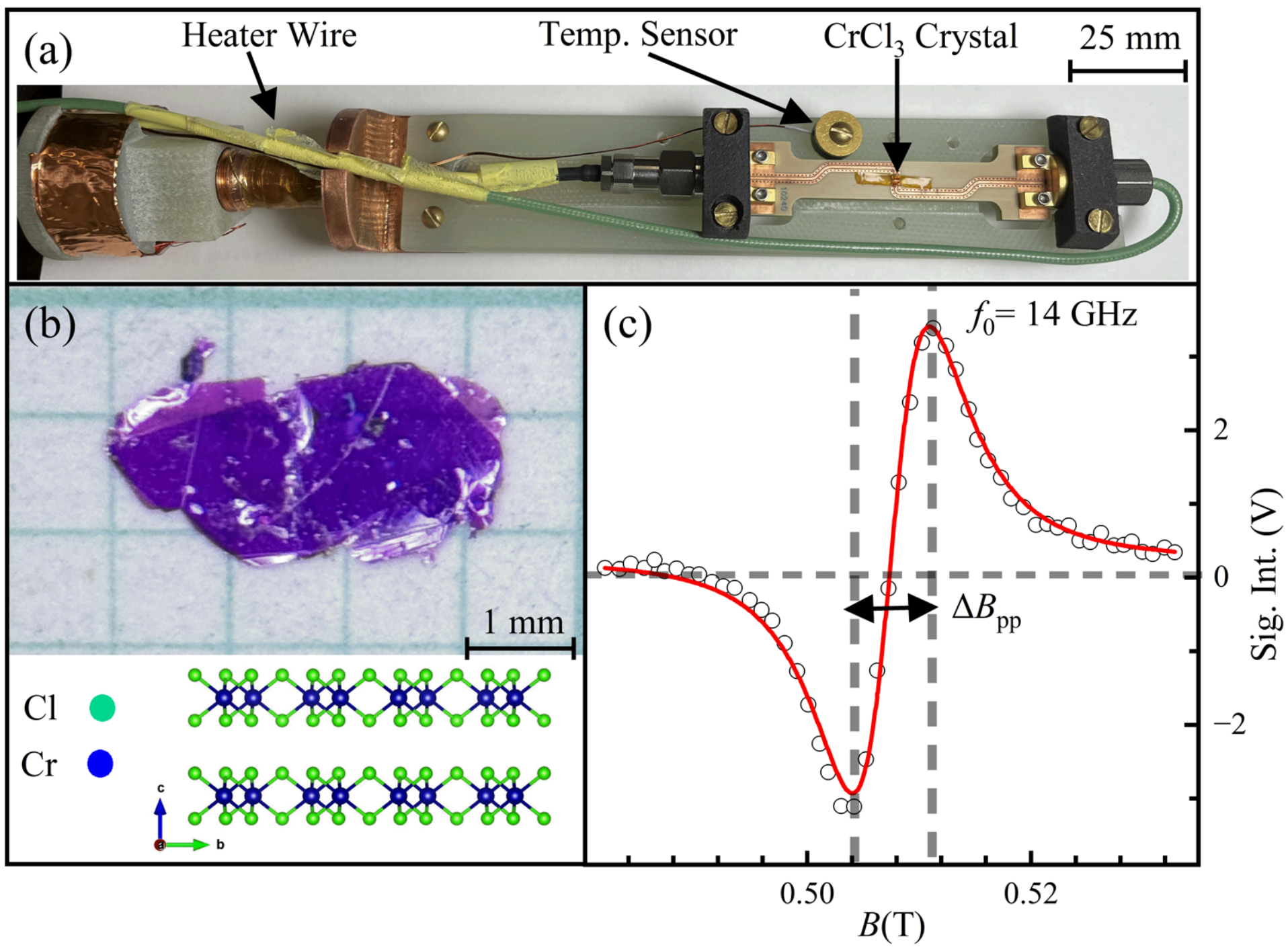

Figure 1. (a) is the picture probe head used for broadband magnetic resonance experiment. The probe head was placed inside the closed cycle cryostat. (b) is the $CrCl_3$ crystal which is used to cleaved to produce uniform $CrCl_3$ sample. (c) is the representative lock-in detected the magnetic resonance spectrum measured at 14 GHz at room temperature. The red-lines is the Lorentzian fit used to extract the peak-peak linewidth.

## Results and Analysis

We measured the magnetic resonance characteristics of $CrCl_3$. Experimentally, we fixed the microwave frequency ($f$) and swept the externally applied magnetic field strength ($B$). We modulated the magnetic field using a pair of modulation coils to enable lock-in detection. Figure 1c shows a representative lock-in-detected magnetic resonance spectrum at 14 GHz at room temperature. The resonance spectrum was fitted with the derivative of a single Lorentzian function [25–27]. From the fit function, we determined the peak-to-peak linewidth ($\Delta B_{pp}$) of the magnetic resonance spectrum. We repeated the measurements at many resonance frequencies between 1 and 16 GHz. Figure 2a shows the measured broadband magnetic resonance spectra at room temperature. Low signal intensity was observed at lower frequencies because spin polarization is low at lower applied magnetic fields. Figure 2b shows overlap of a low-frequency spectrum (3 GHz) with a high-frequency spectrum (16 GHz) for comparison. We observed that the linewidth at low resonance frequency is wider than that at high resonance frequency. We then fit each spectrum using a derivative Lorentzian fit function and plot the linewidth as a function of resonance frequency, as shown in Figure 2c. The linewidth clearly narrows with increasing resonance field. We repeated the measurements from room temperature to 50 K and observed a similar pattern, as shown in the inset of Figure 2c.

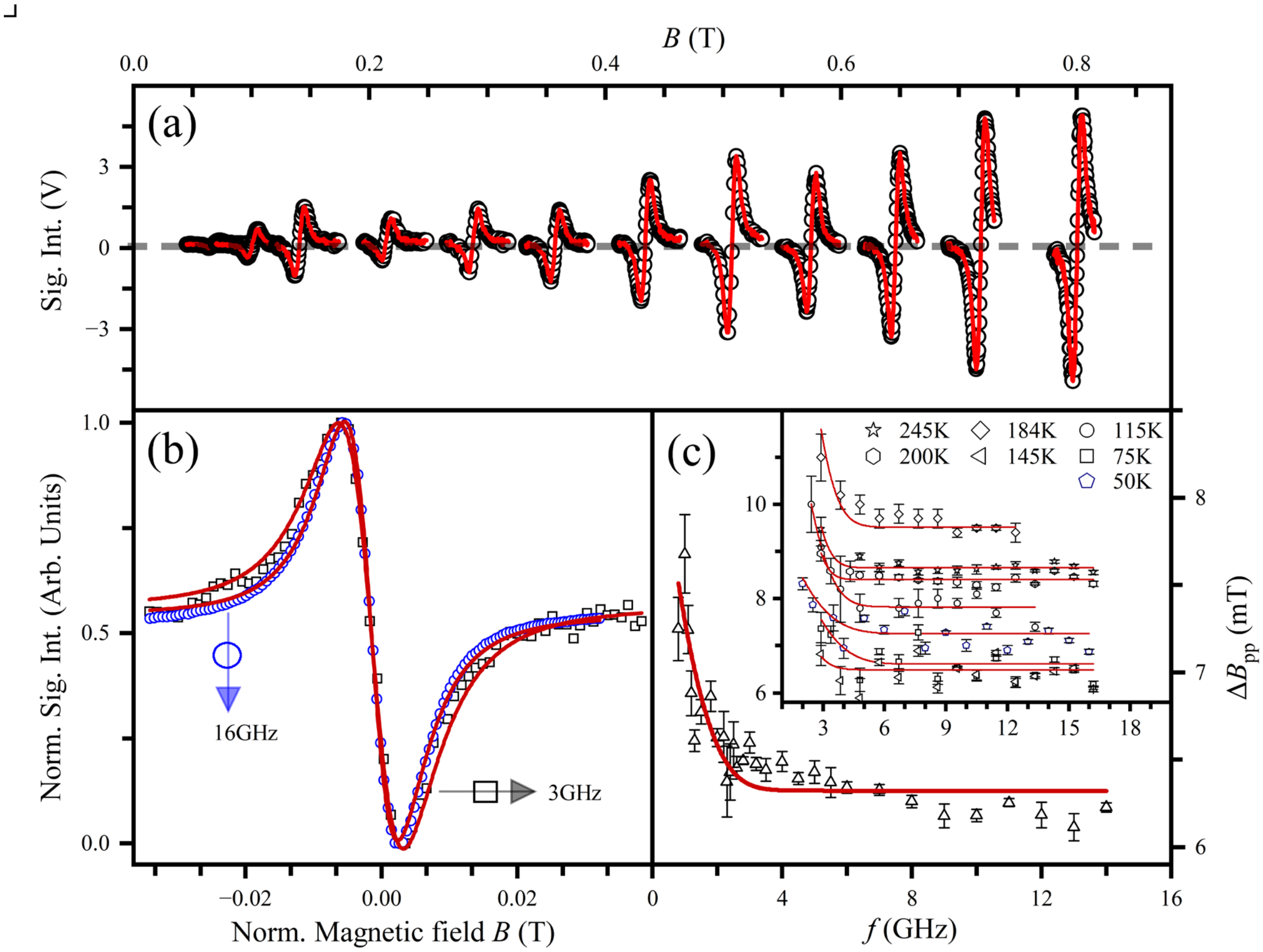

Figure 2 (a) Broadband magnetic resonance spectra of $CrCl_3$ measured over the frequency range from 1 to 16 GHz at room temperature. The red curve is the derivative Lorentzian fits used to estimate the resonance linewidth. (b) Representative resonance spectra measured at 3 GHz and 16 GHz overlap to show the broadening at room temperature (c) Peak-to-peak linewidth $\Delta B_{pp}$ as a function of resonance frequency measured at room temperature demonstrating exchange narrowing of the resonance linewidth. The red line is the fit according to the Anderson–Weiss theory of exchange-narrowing. The inset shows resonance linewidth dependence as function of resonance frequency at different temperatures. Solid red lines represent fits based on the Anderson–Weiss theory of exchange-narrowing.

The observation of spin-resonance linewidth narrowing is a well-established phenomenon in high-spin-density paramagnetic samples with strong spin-exchange interactions. It is also well established that spin exchange leads to a narrowing of the linewidth in the electron spin resonance signal. The theoretical model of exchange narrowing in the electron magnetic resonance signal was proposed by Anderson and Weiss [19,28,29]. The basic idea is that magnetic dipolar interactions between neighboring spins, along with other local interactions, broaden the spin-resonance linewidth. The Anderson-Weiss model predicts that unpaired electrons hop or exchange spin states rapidly with their neighbors when wavefunctions overlap, causing rapid fluctuations in the local magnetic field and averaging out the broadening. As a result, the resonance line shape transforms from a broad Gaussian-like line shape caused by the local magnetic dipole and other fields into a narrow single Lorentzian line shape. This model also predicts that when the resonance frequency ($\omega_0$) is in the same range as the exchange frequency ($\omega_e = \gamma B_{ex}$), the peak-to-peak linewidth ($\Delta B_{pp}$) is proportional to $B_d^2/B_{ex}$, where $B_d$ is the magnetic dipolar field experienced by each spin from all its neighboring spins due to magnetic dipole-dipole interaction and $B_{ex}$ is the exchange field. Another important prediction of the Anderson-Weiss model is that the resonance linewidth becomes independent of the applied field when the applied magnetic field strength is significantly larger than the exchange field. Therefore, as per this model described in the literature [28], the resonance linewidth of electron magnetic resonance is related to various parameters as $\Delta B_{\text{pp}} = \frac{B_d^2}{B_{\text{ex}}}\left\{1 + \frac{5}{3}\mathrm{e}^{-0.5\left(\frac{\omega_0}{\omega_{\text{ex}}}\right)^2} + \frac{2}{3}\mathrm{e}^{-2\left(\frac{\omega_0}{\omega_{\text{ex}}}\right)^2}\right\}$, where $\Delta B_{\text{pp}}$ is the peak-to-peak linewidth measured from the derivative of electron spin resonance spectrum. $B_d$ is the dipolar field, $B_{ex}$ is the exchange field, $\omega_0$ is the resonance frequency, and $\omega_e$ is the exchange frequency.

Anderson-Weiss theory predictions of the exchange process could differ in one- or two-dimensional systems, as in 2D vdW magnets. However, the $CrCl_3$ sample used in the experiment was a macroscopic/bulk sample. Therefore, we concluded that the Anderson–Weiss model was appropriate for interpreting the experimental observations of multifrequency electron magnetic resonance linewidth measurements on a single-crystal bulk $CrCl_3$ sample. As shown in Figure 2c, the multifrequency electron magnetic resonance linewidth was fitted with the Anderson–Weiss model. The model fits the experimental resonance linewidth reasonably well. We repeated the process at all temperatures, as shown in the inset of Figure 2c, from room temperature to 50 K. Below 50 K, the linewidth dependence on the applied magnetic field vanishes due to strong exchange interaction (discussed later). From the fit, we have extracted the exchange ($B_{ex}$) and dipolar ($B_d$) interaction fields at each temperature and plot them in Figure 3.

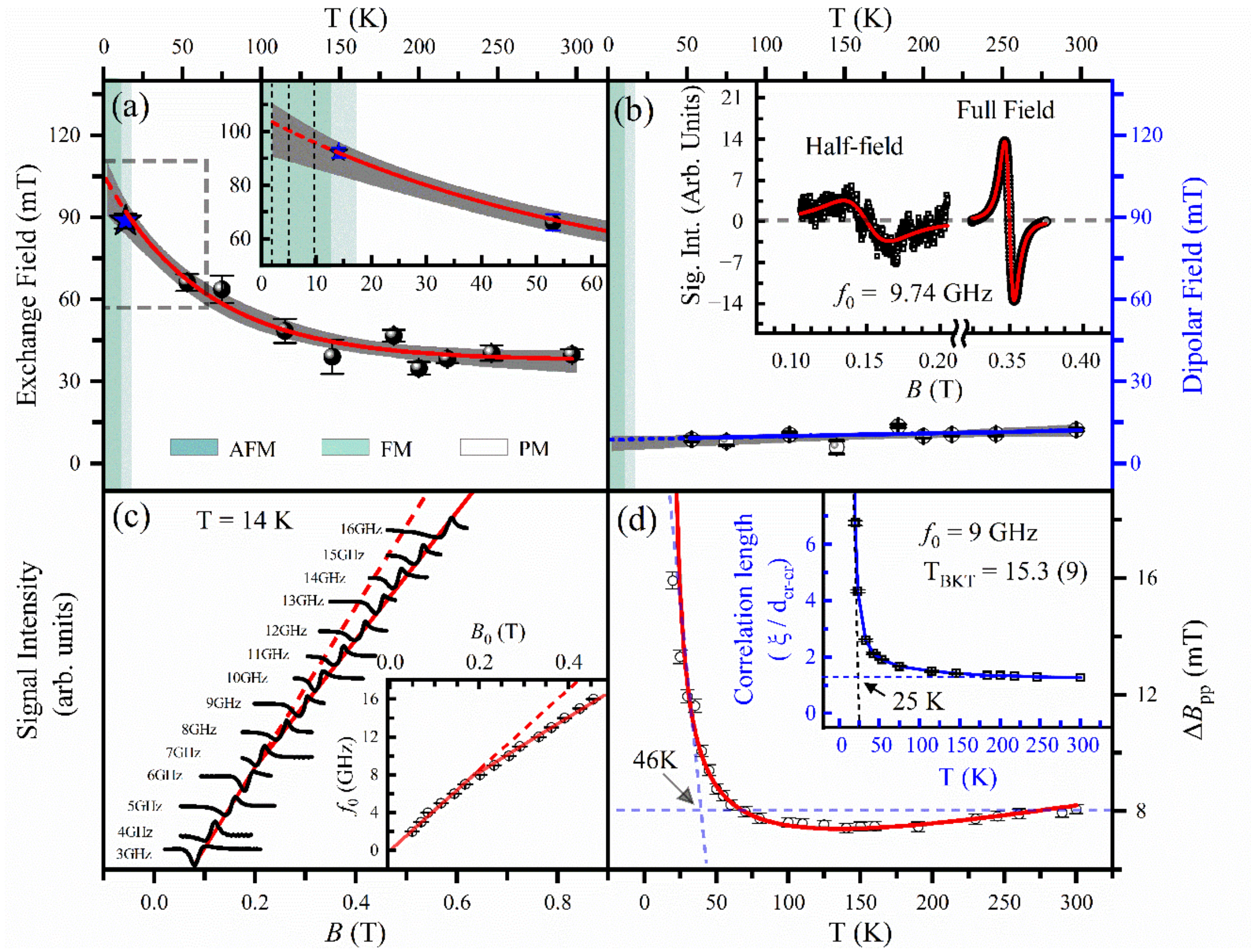

Figure 3 (a) Temperature dependence of the exchange field ($B_{ex}$) (solid circles). The red line is an exponential fit. The gray background region shows the 95% confidence interval calculated using residual bootstrap analysis. The star data point shows the exchange field measured at 14 K. The inset shows the zoomed data shown in the dotted box. The sub-figure in (a) shows the extrapolated exchange interaction up to 1 K. The black dashed lines in the sub-figure in (a) indicate the predicted value of the exchange field at 1.5 K, 5 K, and 10 K. (b) Temperature dependence of dipolar field ($B_d$) (open circles). The solid blue line is the guide to the eye that has been extrapolated to lower temperature. The gray background is a 95% confidence interval obtained from bootstrap analysis. The sub-figure shows the full-field and half-field electron magnetic resonance measurements. The different background colors in (a) and (b) mark the temperature ranges of the different magnetic phases of $CrCl_3$ for clarity. (c) Electron magnetic resonance measurement at 14 K. The sub-figure in (c) shows the resonance field vs resonance frequency plot. The solid red lines represent fits to the resonance modes used to determine the exchange field at 14 K (ferromagnetic regime). (d) shows the electron magnetic resonance linewidth dependence with temperature. The red line represents the fit according to the BKT model. The inset shows the calculated spin correlation length in units of the nearest Cr-Cr distance ($d_{Cr-Cr}$).

Figure 3a shows how the exchange interaction evolves as the temperature decreases. The exchange field shows an exponential enhancement at lower temperatures, indicating reduced thermal/phonon-mediated spin fluctuations. We fit the exchange interactions with an exponential fit shown as a red line in Figure 3a.

We then extrapolated the fit from 50 K to 1 K to estimate the exchange interaction in the ferromagnetic and antiferromagnetic magnetic phases of $CrCl_3$. We used the bootstrap method [16,30] to estimate uncertainty in the predicted values. In brief, we generated randomized data sets using the fit results, with the same noise level as the original data. We then refit these generated data sets with the model, yielding a value for the spread of the fit parameters. A detailed description is provided in the supporting information. The gray background behind the red fit line shows the uncertainty (spread) of the predicted value. Our predicted exchange interaction values were 92(4) mT, 96(3) mT, 100(5) mT, and 105(10) mT at 14 K, 10 K, 5 K, and 1.5 K, respectively, as shown in the subfigure in Figure 3a. To validate the predicted value at 14 K, we performed broadband magnetic resonance measurements at 14 K, as shown in Figure 3c. The subfigure in Figure 3c shows the resonance frequency ($f_0$) plotted as a function of resonance field ($B_0$). Two resonance modes are clearly observed: a high-field uniform mode above approximately 200 mT and a low-field acoustic mode below this value. Similar observations have been reported previously in $CrCl_3$ [31,32]. We fitted the high-field branch using the Kittel equation [33,34] for an in-plane sample in the uniform-precession mode to determine the saturation magnetization. We then used the extracted saturation magnetization to analyze the lower-field acoustic branch and determine the exchange interaction field. From this analysis, the measured exchange field at 14 K is $B_{ex}$=92(4) mT, as shown in Figure 3a (star data point). The measured value agrees well with the value predicted from the multifrequency electron magnetic resonance measurement data from RT to 50 K and with the literature value [7]. We could not measure below 14 K because of limitations in the experimental setup. Therefore, we compared our predicted values with literature values at 1.5 K, 5 K, and 10 K, and found that our predictions agree well with the literature [7,12,32].

Figure 3b shows the evolution of the magnetic dipolar interaction field from room temperature to 50 K. We measured the magnetic dipolar interaction at room temperature to be about 12 mT, which is reasonably strong among paramagnetic spins, indicating that the spins in $CrCl_3$ are strongly coupled via long-range dipolar interactions and supporting the observation that magnetic dipole-dipole interaction broadens the resonance linewidth. We also observed that the magnetic dipolar interaction is nearly independent of the temperature within the experimental error from room temperature to 50 K. The nearly constant magnetic dipole interaction indicates that the relative orientation of the inter-spin vector of the magnetic dipoles does not change with decreasing temperature, indicating the possibility of local/short-range magnetic order. To investigate further, we measured the half-field electron magnetic resonance of $CrCl_3$ at room temperature to estimate the inter-spin distance. A half-field resonance in electron magnetic resonance is a forbidden ($\Delta m_s = \pm 2$) transition occurring at roughly half the applied magnetic field strength of the main resonance line, driven by state mixing in systems with multiple unpaired electrons. Normally, allowed electron spin resonance transitions follow ($\Delta m_s = \pm 1$). In systems with a spin quantum number greater than one ($S \geq 1$), simultaneous changes of two units ($\Delta m_s = \pm 2$) can weakly absorb the microwave energy due to mixing of the pure spin state [35,36]. Spin mixing requires a non-zero zero-field splitting (ZFS) parameter known as the "*D* parameter". The *D* parameter represents the axial zero-field splitting constant governed by the local trigonal crystal-field distortion around the $Cr^{3+}$ ion. The half-field measurement results are shown in the subfigure in Figure 3b. The Black square is the full-field, and the black circle is the half-field electron magnetic resonance measured on the same $CrCl_3$ sample at room temperature. The resonance frequency was 9.74 GHz. The full-field resonance occurred at a $B_0$=350 mT field, and the half-field resonance occurred at 150 mT. From the half- and full-field resonances, we calculated the *D*-parameter to be 0.2(1) $cm^{-1}$, in agreement with the literature value [37]. We utilized the method described by Eaton et *al*. [36] and estimate that the average separation between the magnetic dipoles at room temperature is about 2.4(3) Å, which agrees with the predicted correlation length at room temperature by Mazitov et *al*. [12] and indicates that spins in two nearest-neighbor Cr atoms are correlated or form a short-range magnetic order at room temperature.

To investigate further the short-range spin correlation or short-range magnetic order in $CrCl_3$, we plot the electron magnetic resonance linewidth as a function of temperature, as shown in Figure 3d. The most striking evidence for the evolution of this short-range order is found in the resonance linewidth. We note that an increase in the linewidth at lower temperatures has been reported previously in X-band and higher magnetic field strengths [21,22]. However, we found no reported measurements of linewidth dependence at lower resonance magnetic fields, where the effect of exchange interaction on the resonance linewidth is most visible. Figure 3d shows the linewidth dependence with temperature measured at a resonance frequency of 9 GHz. We also measured the linewidth dependence at many resonance frequencies, both low and high resonance fields, as shown in Figure S6. Figure 3d shows that the linewidth is nearly constant above 250 K, decreases slightly from 250–100 K, and then increases below 100 K. At temperatures lower than 100 K, the average exchange field increases exponentially and becomes larger than the dipolar field, so the exchange interaction dominates the linewidth. The linewidth increase reflects the rapid growth of the magnetic correlation length as the system approaches a more ordered state. Since $CrCl_3$ exhibits easy-plane anisotropy, we fit the temperature-dependent linewidths with the Berezinskii-Kosterlitz-Thouless (BKT) model [38,39]. The red line shows the fit to the experimental data. The fit yields a BKT transition temperature of 15.3(9) K, which agrees with the literature value [12]. As shown in Figure 3d, the tangents of the linewidths at the lower temperature intersect with the constant linewidth at high temperature at about 46 K, indicating a temperature beyond which the magnetic order expands more than the nearest-neighbor distance of Cr atoms and starts the formation of localized ferromagnetic clusters. Using the fit parameters in Figure 3d, we estimate the correlation length in units of the Cr- Cr distance ($d_{Cr\text{-}Cr}$), as described in the literature [12,40,41]. We estimate the correlation length from room temperature to 20 K, as shown in the subfigures in Figure 3d. We observed that as the temperature decreases, the correlation length increases and diverges at about 25 K, which agrees with the PM-to-FM transition temperature for $CrCl_3$. [42–44]

## Discussion and Conclusion

We have experimentally studied the evolution of exchange and magnetic dipolar interactions in $CrCl_3$. The central finding is that the exchange interaction grows exponentially, whereas the dipolar interaction remains unchanged as the temperature decreases. The observation of half-field electron magnetic resonance at room temperature indicates a strong dipolar interaction among the spins, and the invariance of the dipolar interaction with temperature indicates that the orientation of the inter-spin vector does not alter with temperature. The agreement between the inter-spin distance measured by half-field electron spin resonance and the correlation length indicates that short-range order persists well into room temperature. The measured correlation length increases and, at approximately 46 K, marks the onset of localized FM cluster formation; at 25 K, it diverges, indicating a PM-to-FM transition. Finally, we demonstrated that multifrequency broadband electron magnetic resonance is a sensitive and useful method for studying and quantifying the strengths of exchange and dipolar interactions, as well as magnetic order, across the magnetic phases in 2D magnets. This method could be expanded to other vdW and emerging magnetic materials to understand microscopic magnetic properties.

## Acknowledgements

**Funding:** We thank the National Science Foundation for support through Award 2328822. A.T. and J.S. would like to thank the Wichita State University Foundation for support.

**Author contributions:** All authors discussed the results and commented on the paper. K.A. conceived the concept and supervised the project. P.B.K. designed and performed the experiments. P.B.K., A.E.A., J.S., A.I., and P. K analyzed the data. K.P. and J.W. synthesized the $CrCl_3$ crystals. All authors contributed to the preparation of the manuscript.

**Competing interests:** The authors declare no competing interests.

**Data and materials availability:**

All data and code needed to evaluate and reproduce the results in the paper are present in the paper and/or the **Supporting Information**. No new materials were generated during this study. All materials used in this study are commercially available.

**Supporting Information**

$CrCl_3$ crystal characterization; Modulation Parameters Selection; Temperature-dependent broadband magnetic resonance spectra; In-plane vs. out-of-plane measurements; Bootstrap Analysis method; Temperature-Dependent Linewidth of Magnetic Resonance Spectra; Multifrequency linewidth dependence and correlation length measurements.

# Supporting information

# Studying the evolution of exchange and dipolar interactions and spin correlation in vdW magnet $CrCl_3$ across magnetic phases

Prem B. Karki,[1] Ayodimeji E. Aregbesola,[1] Karishma Prasad,[2] Jonathan Skeels,[1] Avani Inamdar[1], Pooja Kalal[1], Jian Wang,[2] Kapildeb Ambal[1]*

[1]Department of Mathematics, Statistics, and Physics, Wichita State University, 1845 Fairmount Street, Wichita, KS 67260, USA

[2]Department of Chemistry and Biochemistry, Wichita State University, 1845 Fairmount Street, Wichita, KS 67260, USA

* Corresponding author: Kapildeb.ambal@wichita.edu

## 1. $CrCl_3$ crystal characterization

Structural characterization was carried out at room temperature on a Rigaku MiniFlex 6G diffractometer employing Cu Kα radiation (λ = 1.5406 Å). Diffraction patterns were recorded between 10° and 80° in 2θ with a step width of 0.05° and a counting time of 10 s per step. Since the material measured was a flake-like single crystal rather than a ground powder, only (00l) reflections appear in the pattern — a signature of strong preferred orientation with the crystallographic c-axis normal to the flake surface. The measured peak positions agree well with a pattern simulated from previously published crystallographic data [1] indicating that the sample is phase pure and adopts the reported structure.

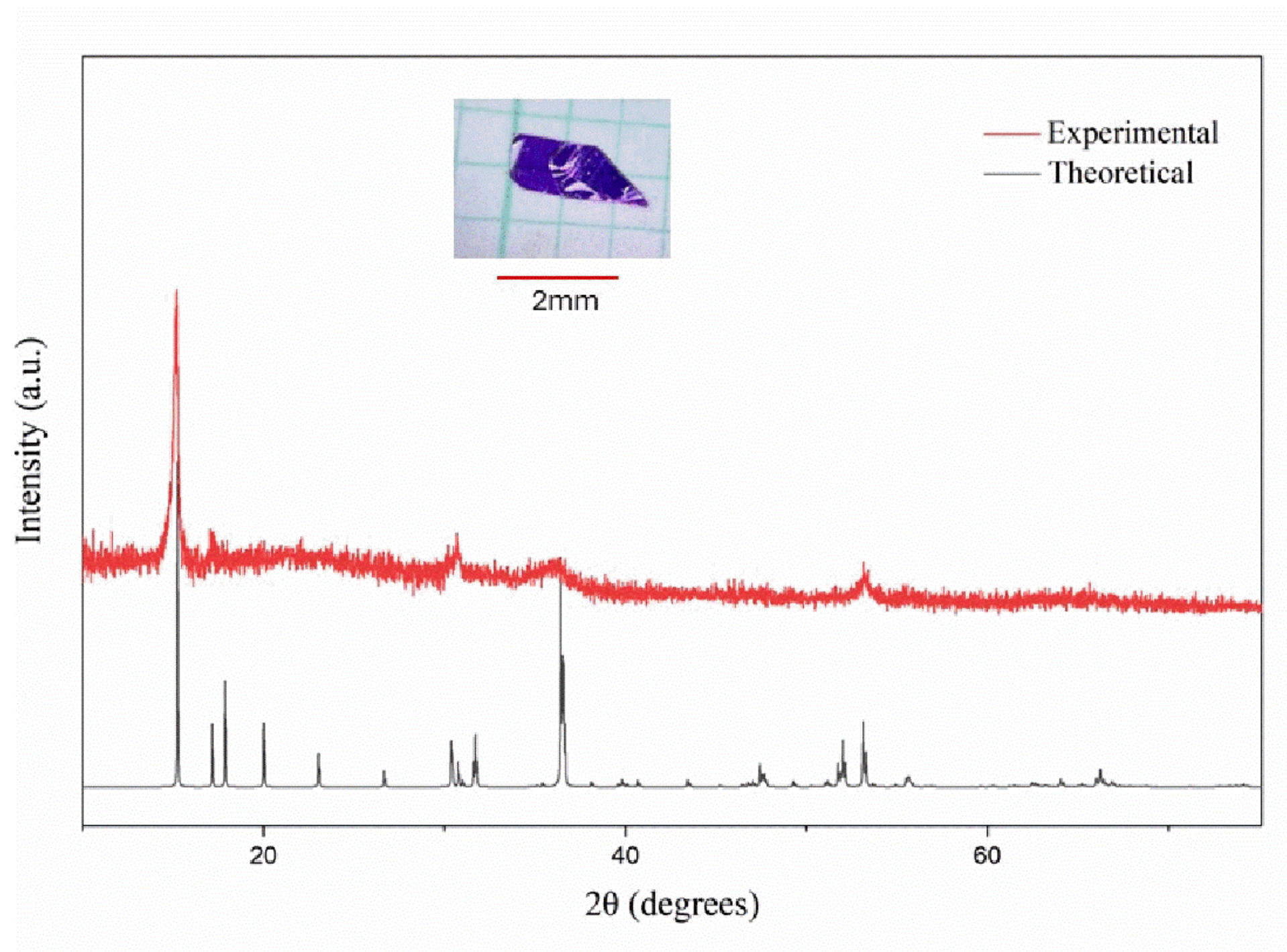

Figure S1: Shows the powder X-ray diffraction pattern of $CrCl_3$. Inset: Optical image of a violet $CrCl_3$ single crystal

## 2. Modulation Parameters selection

Magnetic field modulation often broadens the resonance linewidth and smears the intrinsic linewidth [2]. We determine the optimal modulation parameters by measuring the resonance linewidth as a function of modulation amplitude and frequency and find that 0.5 mT magnetic-field modulation at 77 Hz yields the intrinsic resonance linewidth.

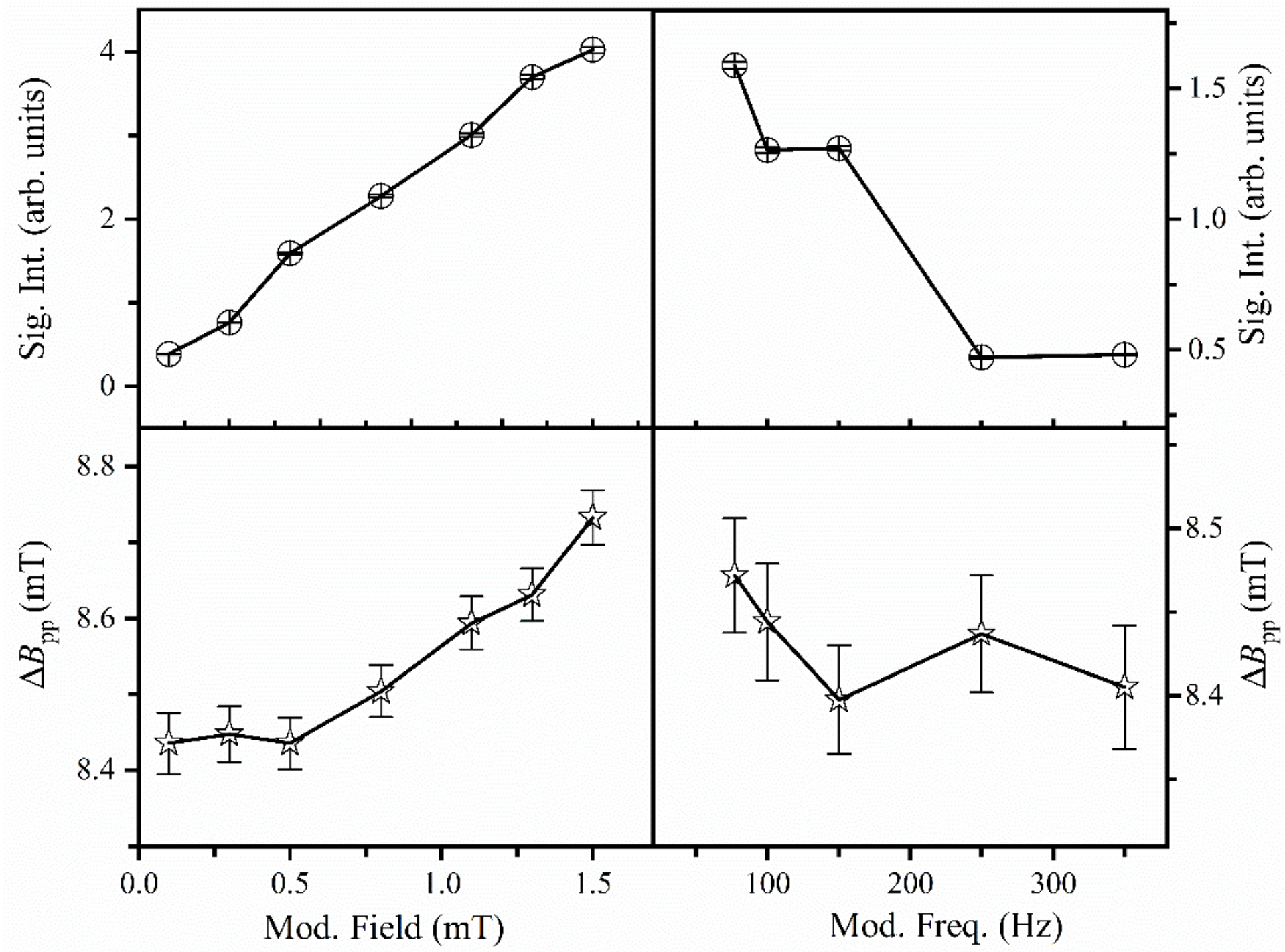

Figure S2 (top row) shows the resonance signal intensity and (bottom row) shows the resonance linewidth measured at 10 GHz as we vary the modulation amplitude and frequency.

### 3. Temperature-dependent broadband magnetic resonance spectra

We acquired broadband magnetic resonance spectra from 1 to 16 GHz over a temperature range of 300 K to 50 K. The solid red lines are the Lorentzian fits of each spectrum.

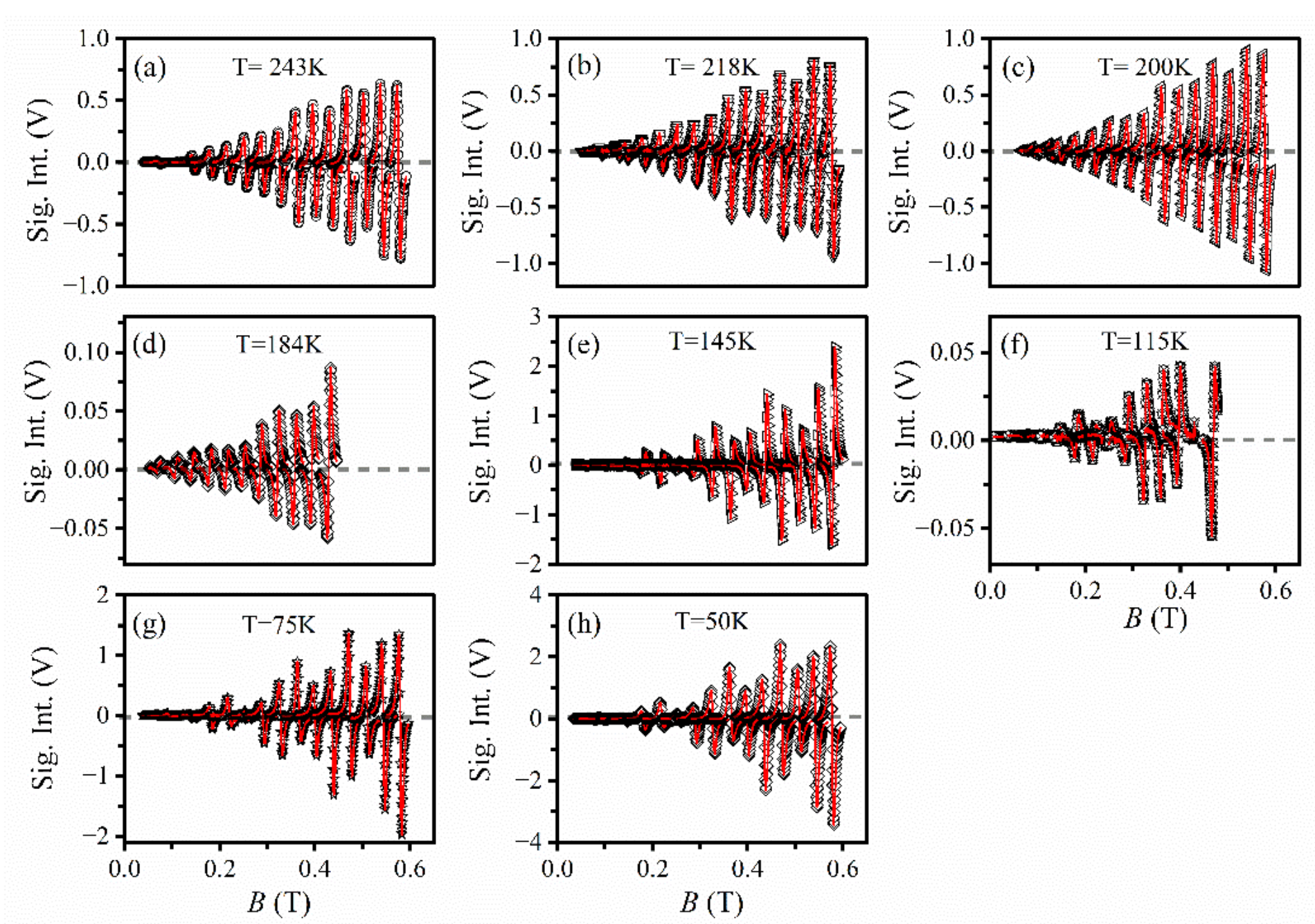

Figure S3. Shows the broadband magnetic resonance spectra from room temperature to 50K.

## 4. In-plane vs. out-of-plane measurements

We measured broadband magnetic resonance in three sample orientations: in-plane (IP), where the applied magnetic field is perpendicular to the c-axis; out-of-plane (OOP), where the applied magnetic field is parallel to the c-axis; and 45 degrees to the c-axis. In all measurements, we ensure the microwave field is perpendicular to the magnetic field, as required for the magnetic resonance condition. Figure S4 (a-c) shows the measured resonance spectra and their dependence on resonance frequencies. We fit each resonance spectrum to extract the resonance field and plotted the results as a function of resonance frequency. Figure S4d shows the plot of resonance field vs. frequency. Using a linear fit, we determined the Landé $g$ factor for each orientation: 1.994(1) for OOP, 1.989(2) for IP, and 1.974(1) for the 45-degree orientation with the c-axis. The measured $g$-factor agrees with the literature values [3–5]. The lower $g$-factor in IP vs OOP indicates $g$-anisotropy, as also reported in the literatures [6]. We also measure the linewidth dependence for these three sample orientations, as shown in Figure S4e. At very low resonance magnetic field (lower resonance frequency), the resonance linewidths overlap for all three orientations. This is because the applied magnetic field is on the order of the internal magnetic field arising from dipole-dipole and exchange interactions. At higher resonance magnetic fields, due to anisotropy, the linewidth in the OOP orientation is larger than in the IP orientation. However, the exchange narrowing effect narrows the linewidth at the same rate for all three orientations, indicating isotropic exchange interaction at room temperature.

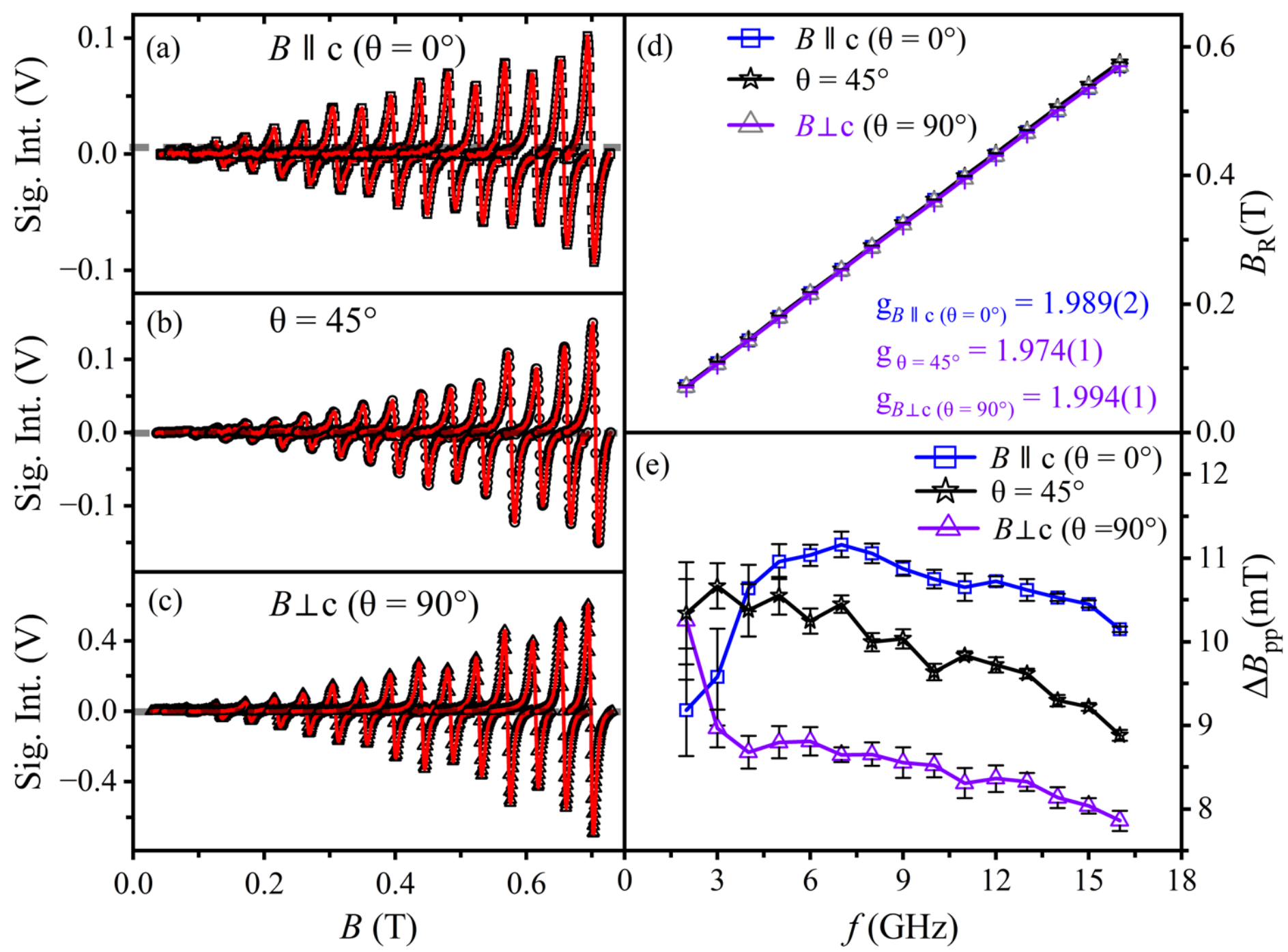

Figure S4. (a-c) shows magnetic resonance spectra at different sample orientations relative to the applied magnetic field strength. Figure S4d shows the resonance field dependence for all three orientations as a function of resonance frequency. Figure S4e shows the linewidth dependence for all three orientations as a function of resonance frequency.

## 5. Bootstrap Analysis method

To evaluate parameter uncertainties and confidence intervals of the fits, we used a residual bootstrap resampling analysis with 10,000 iterations. For the exchange field, we first fitted the experimental data to an exponential function by least-squares regression to establish the baseline fit parameters. We randomly resampled the residuals representing experimental noise with replacement and added them back to the baseline fit to generate synthetic datasets. Refitting these datasets yielded empirical parameter distributions, from which we extracted 95% confidence intervals (represented by the gray regions in Figure 3a). We applied an analogous procedure to the dipolar field measurements using a linear equation. The results are shown as gray regions in Figure 3b.

## 6. Temperature-Dependent Linewidth of Magnetic Resonance Spectra

Figure S5 shows the normalized magnetic resonance spectra measured from room temperature to 20 K. We fitted each spectrum with a Lorentzian function, as described in the main text, to determine the linewidth. Figure S5 shows the broadening of resonance linewidths as a function of temperature.

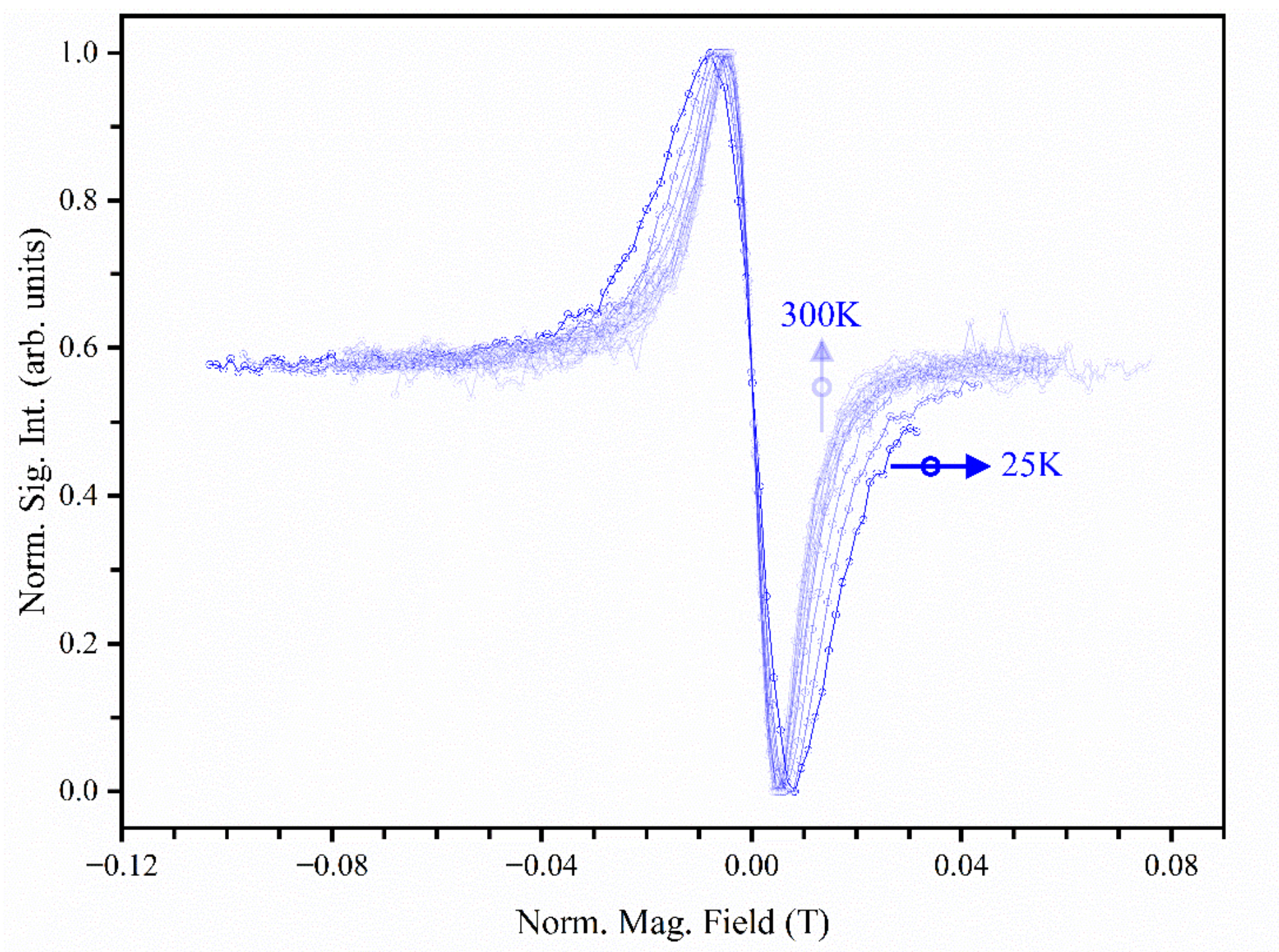

Figure S5. Shows the broadening of resonance linewidths as a function of temperature.

## 7. Multifrequency linewidth dependence and correlation length measurements.

We measured the linewidth and correlation length at various resonance fields, both at low resonance fields ($f_0$ = 5 GHz and $f_0$ = 7 GHz), where dipolar fields predominantly dominate the linewidth, and at high fields ($f_0$ = 9.75 GHz and $f_0$ = 15 GHz), where the resonance linewidth is narrowed due to exchange narrowing. We fit all temperature-dependent linewidth measurements with the BKT model (solid red lines in Figure S6), yielding a BKT temperature of about 15(1) K. We estimate the correlation length, which is shown as an inset in the respective Figures. In all resonance fields, the correlation length diverges around 25 K, which agrees with the PM-to-FM transition temperature for $CrCl_3$.

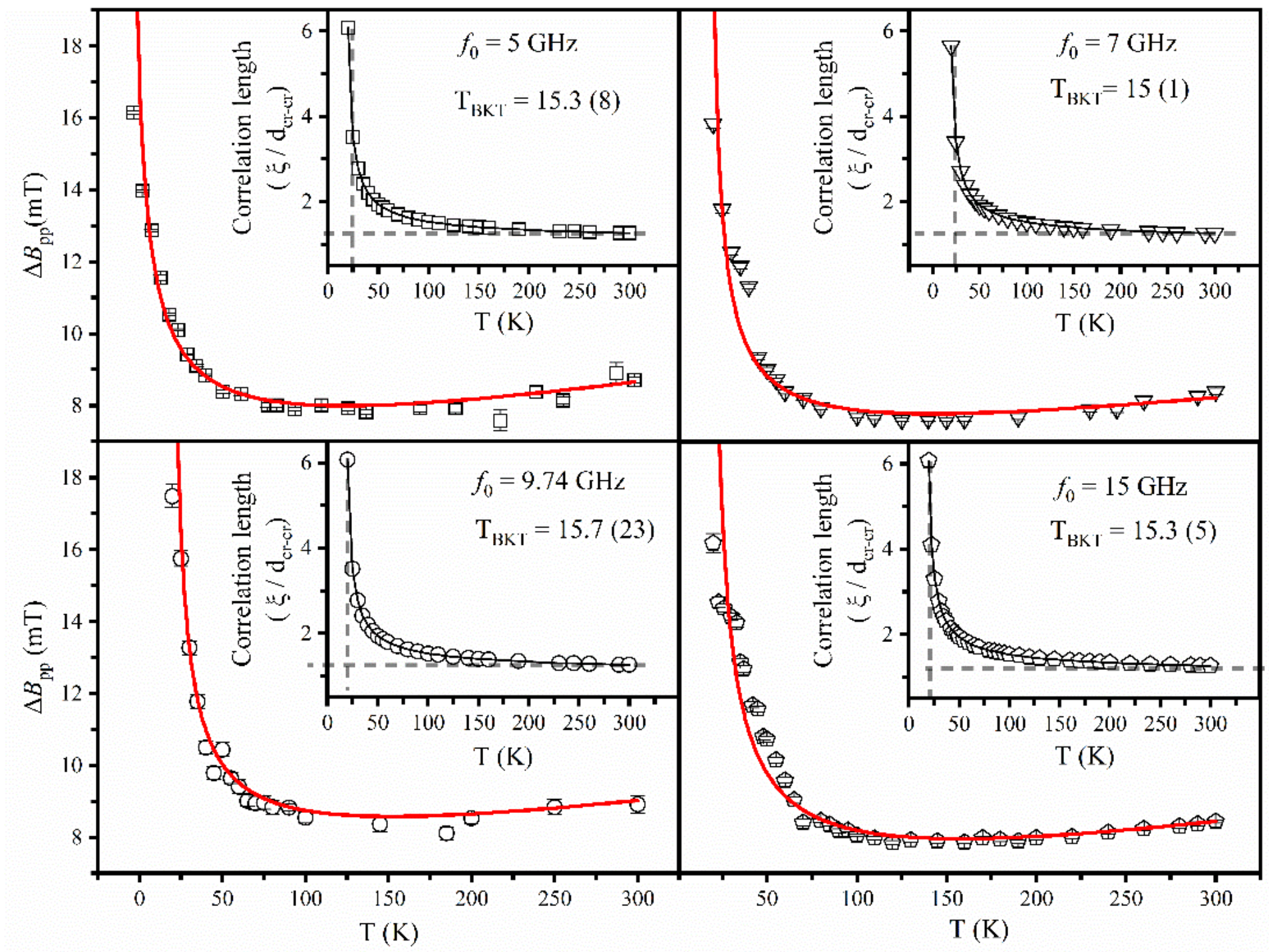

Figure S6. Shows the multifrequency resonance linewidth dependence on temperature. The solid red line is the fit using the BKT model. The inset in each figure is the correlation length as a function of temperature.